\documentclass[11pt]{article}

\usepackage[letterpaper, margin=1.2in]{geometry}
\usepackage[T1]{fontenc}
\usepackage[english]{babel}

\usepackage{graphicx}
\usepackage{booktabs}
\usepackage{array}
\usepackage{longtable}
\usepackage{float}               
\usepackage{caption}
\usepackage{titlesec}
\titleformat{\section}{\normalfont\Large\bfseries}{\thesection.}{0.6em}{}
\titleformat{\subsection}{\normalfont\large\bfseries}{\thesubsection.}{0.6em}{}
\titlespacing*{\section}{0pt}{1.6em}{0.8em}
\titlespacing*{\subsection}{0pt}{1.2em}{0.6em}

\usepackage{setspace}
\usepackage[hidelinks]{hyperref}
\usepackage{enumitem}
\begin{document}

\begin{center}
{\Large\bfseries BLUE: Semantics-Preserving Video Compression for Efficient\\[2pt]
Vision-Language Surveillance Analytics}

\vspace{1.4em}
{\large Shubham Baid}

\vspace{0.6em}
{\large Akash James}

\vspace{0.6em}
{\large Sahil Chachra}

\vspace{0.6em}
{\large Nishant Sinha}

\vspace{0.6em}
{\large Kunal Kislay}

\vspace{1em}
{\large KGraph AI Solutions Pvt.\ Ltd., Bangalore, India -- 560016}
\end{center}

\vspace{1.5em}

\begin{center}\bfseries Abstract\end{center}
\noindent
Continuous surveillance video creates a growing storage, transmission, and inference burden for
enterprise video analytics systems. While modern codecs such as H.265 reduce bitrate for
human-viewable video, aggressive compression can degrade downstream computer-vision performance
and does not necessarily reduce the number of vision-language model (VLM) inference calls required
for semantic video understanding. This paper evaluates BLUE, a fixed-camera surveillance
compression approach that suppresses static-background redundancy while preserving foreground
activity, for its effect on VLM-based event and anomaly understanding. We compare raw H.265 and
BLUE-compressed H.265 video on two surveillance datasets: VIRAT, comprising 227 paired event
samples from 106 clips, and CHAD, comprising 54 human-activity anomaly clips. For each pair, the
same frame index is evaluated using a VLM captioning pipeline, and outputs are scored against
annotation-derived ground truth using a blind judging protocol. The results show no measurable
degradation in semantic inference quality. On VIRAT, the mean VLM score remains effectively
unchanged between raw H.265 and BLUE, with a mean difference of approximately -0.01 on a 0-10
scale. On CHAD, raw H.265 and BLUE obtain near-equivalent mean scores of 4.31 and 4.26,
respectively. Compression saving is also uncorrelated with VLM score change on VIRAT ($r = 0.004$),
indicating that higher BLUE compression does not predict semantic quality loss. Beyond storage
reduction, BLUE increases the share of skip-heavy P-frames on CHAD from 1.4\% to 53.2\%, enabling
an estimated 53\% reduction in VLM calls through packet-size-based frame skipping. These findings
suggest that BLUE functions as a machine-centric compression layer for surveillance video, reducing
bandwidth and inference cost while preserving VLM semantic performance.

\vspace{0.4em}

\section{Introduction}

Fixed-camera surveillance deployments are expanding across enterprises, public infrastructure,
logistics facilities, manufacturing plants, retail outlets, and transportation systems. These
deployments increasingly require video to be both archived efficiently and analyzed automatically.
The emergence of vision-language models (VLMs) further changes the role of surveillance video:
instead of only supporting human review or closed-set object detection, video streams can now be
queried and summarized in natural language. However, continuous high-resolution video from large
camera networks creates substantial storage, transmission, and inference costs.

Modern video codecs such as H.264/AVC, H.265/HEVC, AV1, and H.266/VVC have achieved substantial
rate-distortion gains, with HEVC and VVC designed to provide major bitrate reductions relative to
earlier standards \cite{ref1,ref2,ref3}. Learned image and video compression methods have also
advanced rapidly \cite{ref4,ref5}. Nevertheless, most mainstream coding approaches remain optimized
primarily for human perceptual reconstruction rather than for downstream machine understanding.

This mismatch has motivated the field of video coding for machines (VCM), where the objective is to
preserve task-relevant information for object detection, segmentation, tracking, and analytics
rather than reconstruct every pixel for human viewing \cite{ref6,ref7}. Prior studies show that
video quality and compression can affect machine-vision performance, including object detection and
tracking, particularly under high compression, small objects, low light, or fast motion
\cite{ref8,ref9,ref10,ref11}. Thus, surveillance systems face a practical tension: they require
aggressive compression, but conventional compression may degrade the visual cues needed by AI
models.

The challenge is different, but equally important, for VLM-based surveillance analysis. VLMs can
describe scenes and reason about anomalies in an open-ended manner, but their inference cost is
driven largely by the number of frames or visual tokens processed. Reducing encoded file size alone
does not necessarily reduce VLM cost after frames are decoded and tokenized. Recent work on
multimodal long-context processing and long-video understanding has similarly emphasized that dense
visual streams create large token and compute burdens, while naive sampling may waste capacity on
redundant frames or miss decisive moments \cite{ref12,ref13}.

BLUE addresses this gap through a fixed-camera, machine-centric compression strategy. It maintains
a persistent background representation and substitutes unchanged static background regions with
seeded background pixels before conventional H.265 encoding, while preserving foreground activity.
This enables the encoder to represent static background regions using skip-heavy or
near-zero-motion blocks, reducing file size without discarding semantically relevant foreground
content. It also produces a useful bitstream-level signal: very small P-frames indicate little new
scene information and can be skipped by downstream VLM pipelines.

This paper evaluates whether BLUE compression affects VLM semantic inference quality on
surveillance video. We compare raw H.265 and BLUE-compressed H.265 variants on two datasets: VIRAT,
containing vehicle and pedestrian events, and CHAD, containing human activity anomalies. The
evaluation uses paired frame selection, VLM captioning, annotation-derived ground truth, and blind
semantic scoring. We further analyze whether BLUE-induced P-frame sparsity can reduce downstream VLM
calls. The results show that BLUE preserves semantic inference quality while reducing file size and
enabling approximately half of VLM calls to be skipped on CHAD.

\section{Methodology}

\subsection{Datasets}

The experiment is designed as a paired comparison between raw H.265 video and BLUE-compressed H.265
video. The same frame index is evaluated in both variants so that score differences arise from the
compression pipeline rather than differences in event timing or scene content. The benchmark uses
VIRAT and CHAD, which provide complementary surveillance settings.

VIRAT is used as the primary event-level benchmark because it contains realistic outdoor
surveillance scenes involving vehicles, pedestrians, and human-object interactions \cite{ref14}. The
VIRAT evaluation consists of 106 clips and 227 paired event samples, with event types including
loading/unloading objects, opening/closing trunks, entering/exiting vehicles, gesturing, carrying
objects, running, and entering/exiting facilities.

CHAD is used as a complementary human-anomaly benchmark \cite{ref15}. The subset used here contains
54 clips at 1920 $\times$ 1080 resolution and 30 fps, with clip durations from 15 to 115 seconds.
CHAD provides frame-level anomaly labels and person-tracking metadata, which are used to select
activity-rich frames and support the P-frame sampling analysis. The category composition of the
CHAD subset is listed in Table \ref{tab:chad-dist}.

\begin{table}[H]
\centering
\caption{CHAD category distribution.}
\label{tab:chad-dist}
\begin{tabular}{lr}
\toprule
\textbf{Category} & \textbf{Clips} \\
\midrule
Normal & 6 \\
Riding a bicycle & 5 \\
Fight & 8 \\
Person lying on floor & 10 \\
Running away & 10 \\
Robbery & 6 \\
Playing & 4 \\
Throwing & 5 \\
\midrule
\textbf{Total} & \textbf{54} \\
\bottomrule
\end{tabular}
\end{table}

\subsection{Video Preparation and Frame Selection}

For each evaluated sample, two variants are generated. The baseline is encoded using libx265 at CRF
28. The BLUE variant applies the BLUE pre-encoding transformation and is then encoded using the
same libx265 CRF 28 setting. Thus, the comparison is between $V_{raw} = H.265(X)$ and $V_{BLUE} =
H.265(BLUE(X))$, where $X$ denotes the original video. Since the final codec and CRF are identical,
the evaluation isolates the effect of BLUE.

The VLM is evaluated on representative frames rather than full video sequences. For VIRAT, one
frame is selected from the annotated event window for each event sample. For CHAD, the frame is
selected from the labeled anomaly window; for normal clips, a representative non-anomalous frame is
selected. In both datasets, the raw and BLUE variants use the same frame index.

\subsection{VLM Inference and Scoring}

The captioning model is Qwen3.6-35B-A3B quantized to 4-bit precision. VIRAT is evaluated using the
AWQ 4-bit model through vLLM \cite{ref16}, while CHAD is evaluated locally using mlx-vlm 0.6.1 on
Apple Silicon. Within each dataset, raw and BLUE frames are processed using the same prompt, model
configuration, and temperature setting. The experimental configuration is summarized in Table
\ref{tab:config}.

\begin{table}[H]
\centering
\caption{Experimental configuration.}
\label{tab:config}
\begin{tabular}{p{2.2cm}p{2.3cm}p{2.3cm}}
\toprule
\textbf{Component} & \textbf{VIRAT} & \textbf{CHAD} \\
\midrule
VLM model & Qwen3.6-35B-A3B-AWQ-4bit & Qwen3.6-35B-A3B-4bit \\
Inference backend & vLLM & mlx-vlm 0.6.1 (Apple Silicon MPS) \\
Raw encoder & libx265 CRF 28 & libx265 CRF 28 \\
BLUE encoder & BLUE pipeline $\rightarrow$ libx265 CRF 28 & BLUE pipeline $\rightarrow$ libx265 CRF 28 \\
Temperature & 0.0 & 0.0 \\
\bottomrule
\end{tabular}
\end{table}

The prompt instructs the VLM to describe the surveillance frame in two to three sentences, focusing
on the people, vehicles, objects, interactions, and any notable activity visible in the frame.
Ground truth is derived from dataset annotations. Generated captions are scored by an independent
judging model, Qwen3.5-122B-A10B, using a blind protocol in which the raw and BLUE captions for each
sample are presented to the judge in randomized, reversed order so that it cannot infer which
variant it is scoring. Scores are assigned on a 0-1 scale and reported on a 0-10 scale. For each
paired sample $i$, the BLUE effect is measured as $\Delta S_i = S_{i,BLUE} - S_{i,raw}$. Positive
values indicate BLUE wins, negative values indicate raw H.265 wins, and zero indicates a tie.

\subsection{Metrics and P-Frame Analysis}

The primary metrics are mean semantic score, mean score difference, win/tie/loss counts, and
category-level deltas. Compression saving is computed as $\mathrm{Saving} = (1 -
\mathrm{Size}_{BLUE} / \mathrm{Size}_{raw}) \times 100$. For VIRAT, Pearson correlation is computed
between compression saving and VLM score change to test whether higher compression predicts
semantic degradation.

For CHAD, P-frame packet sizes are extracted from the H.265 bitstream using ffprobe \cite{ref17}. A
P-frame below 5 KB is treated as skip-heavy, indicating little new scene information. The estimated
VLM call rate is computed as the fraction of frames above this threshold. This evaluates whether
BLUE provides a low-cost bitstream signal for skipping redundant frames before VLM inference.

\section{Results and Discussion}

\subsection{Overall VLM Inference Quality}

Table \ref{tab:overall} summarizes the main result. Across both VIRAT and CHAD, BLUE compression
produces no measurable degradation in VLM semantic inference quality relative to raw H.265. On
VIRAT, the mean score is effectively unchanged at 4.59/10 for both variants, with a mean difference
of approximately -0.01. On CHAD, the mean score changes only slightly from 4.31/10 for raw H.265 to
4.26/10 for BLUE.

\begin{table}[H]
\centering
\caption{Overall VLM inference quality for raw H.265 and BLUE.}
\label{tab:overall}
\resizebox{\linewidth}{!}{%
\begin{tabular}{lccccccc}
\toprule
\textbf{Dataset} & \textbf{Samples} & \textbf{Raw mean} & \textbf{BLUE mean} & \textbf{Mean $\Delta$} & \textbf{BLUE wins} & \textbf{Ties} & \textbf{BLUE losses / File-size reduction} \\
\midrule
VIRAT & 227 event samples & 4.59 & 4.59 & -0.01 & 55 & 116 & 56 / 60--84\% \\
CHAD & 54 clips & 4.31 & 4.26 & -0.06 & 13 & 30 & 11 / 40--55\% \\
\bottomrule
\end{tabular}}
\end{table}

\begin{figure}[H]
\centering
\includegraphics[width=0.85\linewidth]{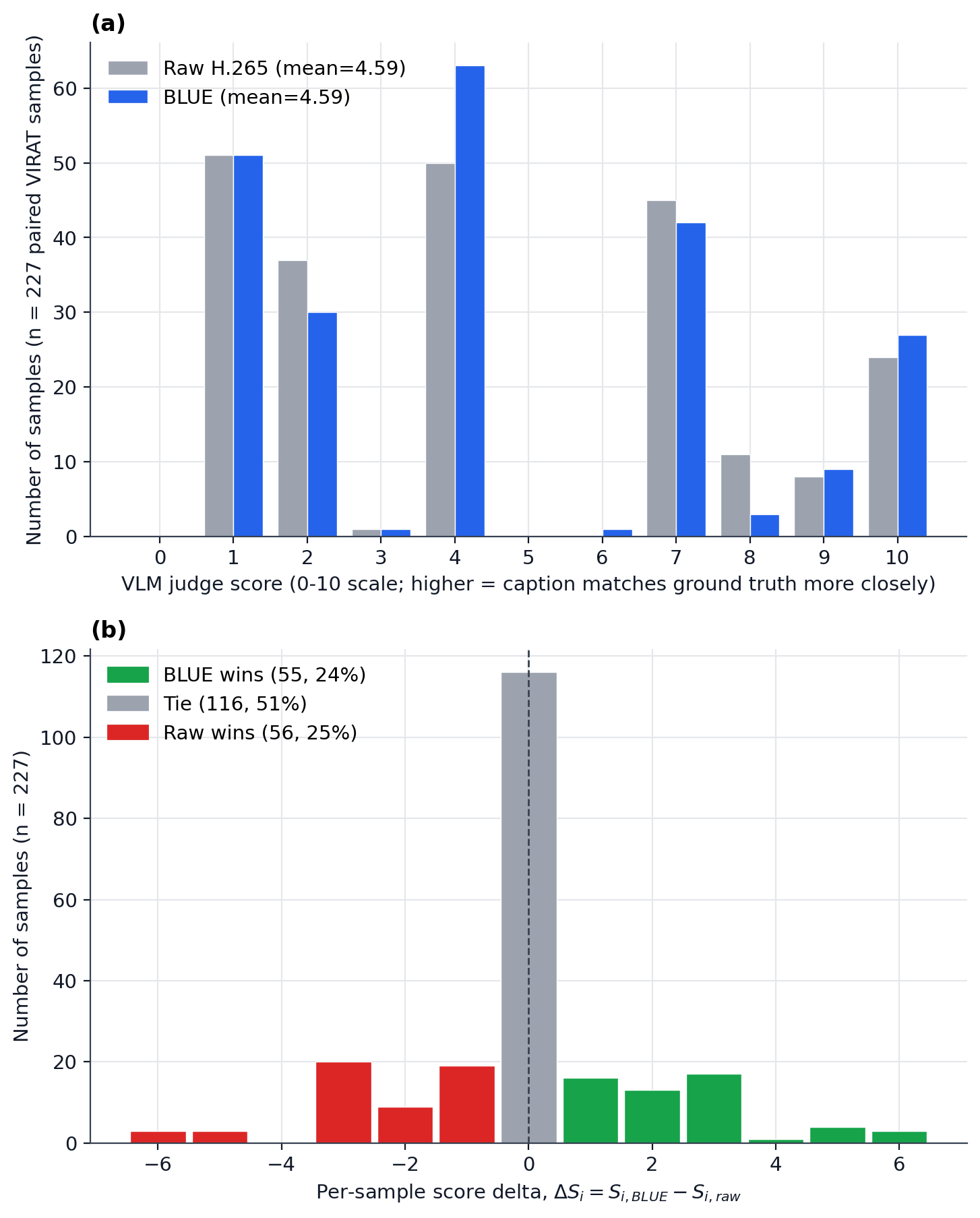}
\caption{VIRAT raw H.265 vs.\ BLUE score comparison ($n = 227$ paired samples). (a) Score
distribution: each bar gives the number of samples at a given VLM judge score (0--10 scale; higher
= generated caption matches ground truth more closely). The distributions overlap almost
completely, with means of 4.59 for both variants. (b) Per-sample score delta, $\Delta S_i =
S_{i,BLUE} - S_{i,raw}$. Green = BLUE scores higher (55 samples, 24\%), grey = tie (116 samples,
51\%), red = raw H.265 scores higher (56 samples, 25\%); the distribution is centered near zero.}
\label{fig:1}
\end{figure}

The paired win/tie/loss counts in Table \ref{tab:overall} are particularly important. VIRAT shows
near-perfect symmetry, with BLUE scoring higher in 55 samples and lower in 56 samples, while 116 are
tied. CHAD also shows no systematic disadvantage for BLUE: most clips are tied and BLUE wins
slightly more clips than raw H.265. As shown in Fig. \ref{fig:1}(a)--(b), the VIRAT raw and BLUE
score distributions overlap almost completely and the per-sample delta is centered near zero,
indicating no leftward shift in semantic quality for BLUE.

\begin{figure}[H]
\centering
\includegraphics[width=\linewidth]{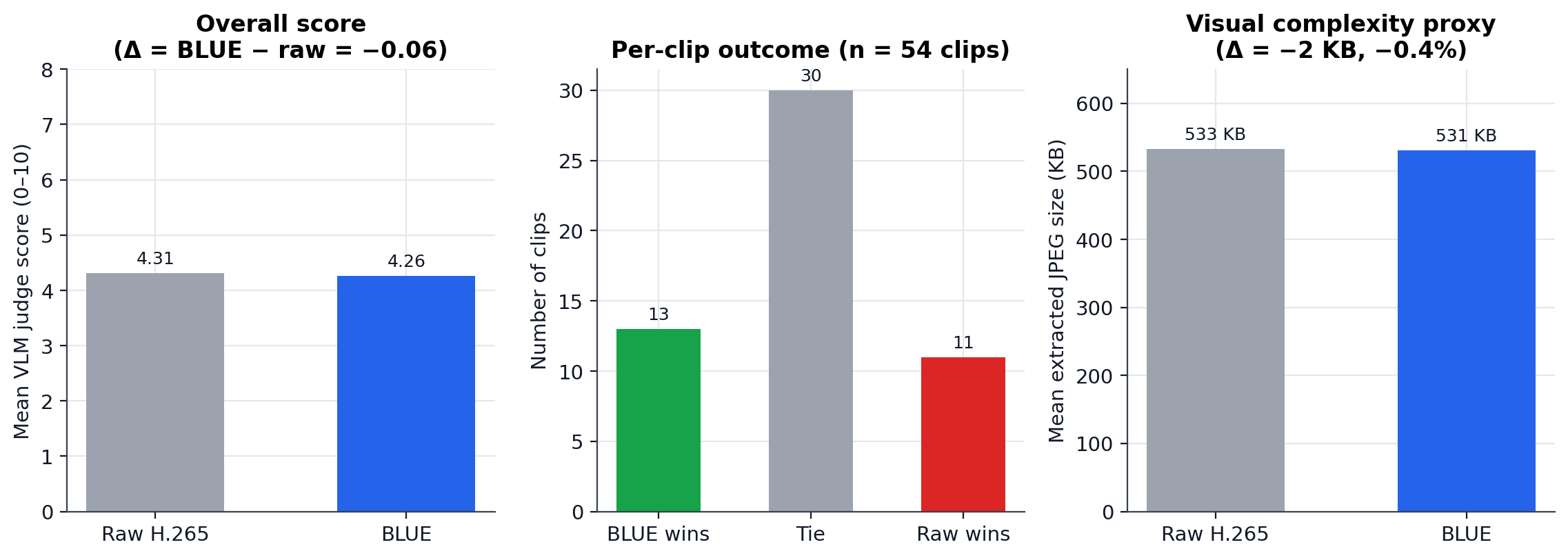}
\caption{CHAD benchmark summary, raw H.265 vs.\ BLUE ($n = 54$ clips). Left: mean VLM judge score
(0--10) per variant, $\Delta$ = BLUE mean $-$ raw mean = $-0.06$. Center: per-clip outcome counts --
clips where BLUE scored higher (``BLUE wins''), the two variants tied, or raw H.265 scored higher
(``Raw wins''). Right: mean extracted JPEG size per frame (KB), used as a proxy for the amount of
visual complexity passed into the VLM; the near-identical values show BLUE does not reduce the
visual information the VLM has to process per frame.}
\label{fig:2}
\end{figure}

Similarly, Fig. \ref{fig:2} shows that the CHAD score distributions have strong overlap across the
score range, further supporting the conclusion that BLUE does not systematically degrade VLM scene
understanding.

\subsection{VIRAT Event-Level Results}

The VIRAT event-level breakdown in Table \ref{tab:event-type} shows how BLUE behaves across
surveillance activity types. Most event categories track closely between raw H.265 and BLUE, with
both positive and negative deviations distributed across event types.

\begin{table}[H]
\centering
\caption{VIRAT event-type score breakdown.}
\label{tab:event-type}
\begin{tabular}{lcccc}
\toprule
\textbf{Event type} & \textbf{Raw} & \textbf{BLUE} & \textbf{$\Delta$} & \textbf{$n$} \\
\midrule
Loading object into vehicle & 0.47 & 0.61 & +0.14 & 12 \\
Unloading from vehicle & 0.52 & 0.52 & 0.00 & 25 \\
Opening trunk & 0.33 & 0.29 & -0.04 & 17 \\
Closing trunk & 0.39 & 0.32 & -0.07 & 16 \\
Getting into vehicle & 0.31 & 0.34 & +0.03 & 39 \\
Getting out of vehicle & 0.51 & 0.48 & -0.03 & 19 \\
Gesturing & 0.36 & 0.44 & +0.08 & 5 \\
Carrying object & 0.58 & 0.56 & -0.02 & 81 \\
Running & 0.13 & 0.15 & +0.02 & 6 \\
Entering facility & 0.40 & 0.55 & +0.15 & 2 \\
Exiting facility & 0.42 & 0.30 & -0.12 & 5 \\
\bottomrule
\end{tabular}
\end{table}

\begin{footnotesize}
$n$ is the number of paired VIRAT event samples in that category, out of 227 total. $\Delta$ =
mean(BLUE) $-$ mean(raw), computed per sample on the 0-1 scale described in Section II-C, before
conversion to the 0-10 scale used elsewhere in this paper; positive values indicate BLUE scoring
higher than raw H.265. Categories with small $n$ (e.g., entering facility, $n = 2$) should be read
as indicative rather than statistically robust.
\end{footnotesize}

\begin{figure}[H]
\centering
\includegraphics[width=\linewidth]{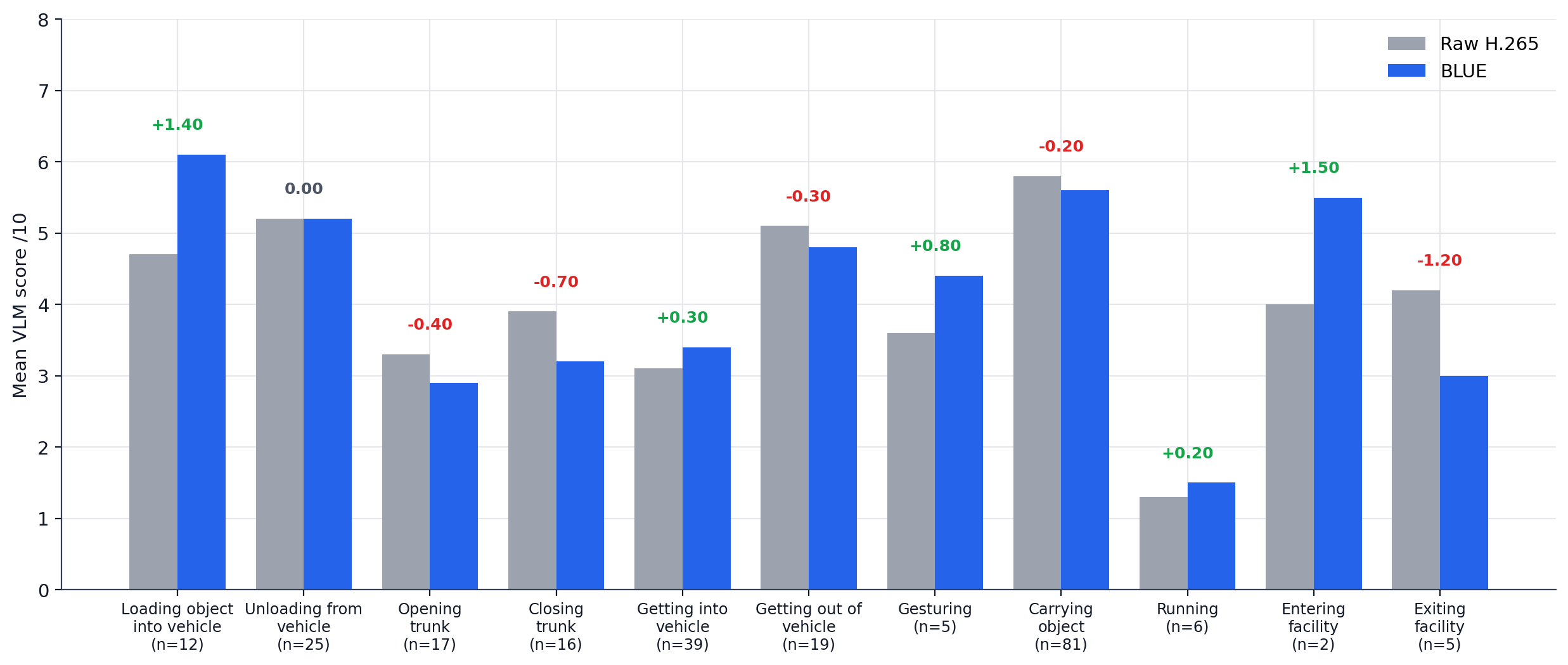}
\caption{Mean VIRAT score by event type, raw H.265 vs.\ BLUE. $n$ (below each category label) is the
number of paired samples of that type, out of 227 total. $\Delta$ (above each bar pair) =
mean(BLUE) $-$ mean(raw) on the 0--10 scale, i.e., the values in Table \ref{tab:event-type}
multiplied by 10 for consistency with the other figures in this paper.}
\label{fig:3}
\end{figure}

As shown in Fig. \ref{fig:3}, raw H.265 and BLUE scores track closely for most VIRAT event types.
The strongest positive deltas occur for entering facility (+0.15), loading object into vehicle
(+0.14), and gesturing (+0.08), although some of these categories have small sample sizes. The
largest high-volume category, carrying object ($n = 81$), is nearly unchanged with a delta of
-0.02. The largest regression occurs for exiting facility (-0.12), followed by closing trunk
(-0.07), suggesting that short interactions near doors, trunks, and static structures may be more
sensitive to transient foreground-background overlap.

\subsection{Compression Saving and Quality Change}

\begin{figure}[H]
\centering
\includegraphics[width=0.85\linewidth]{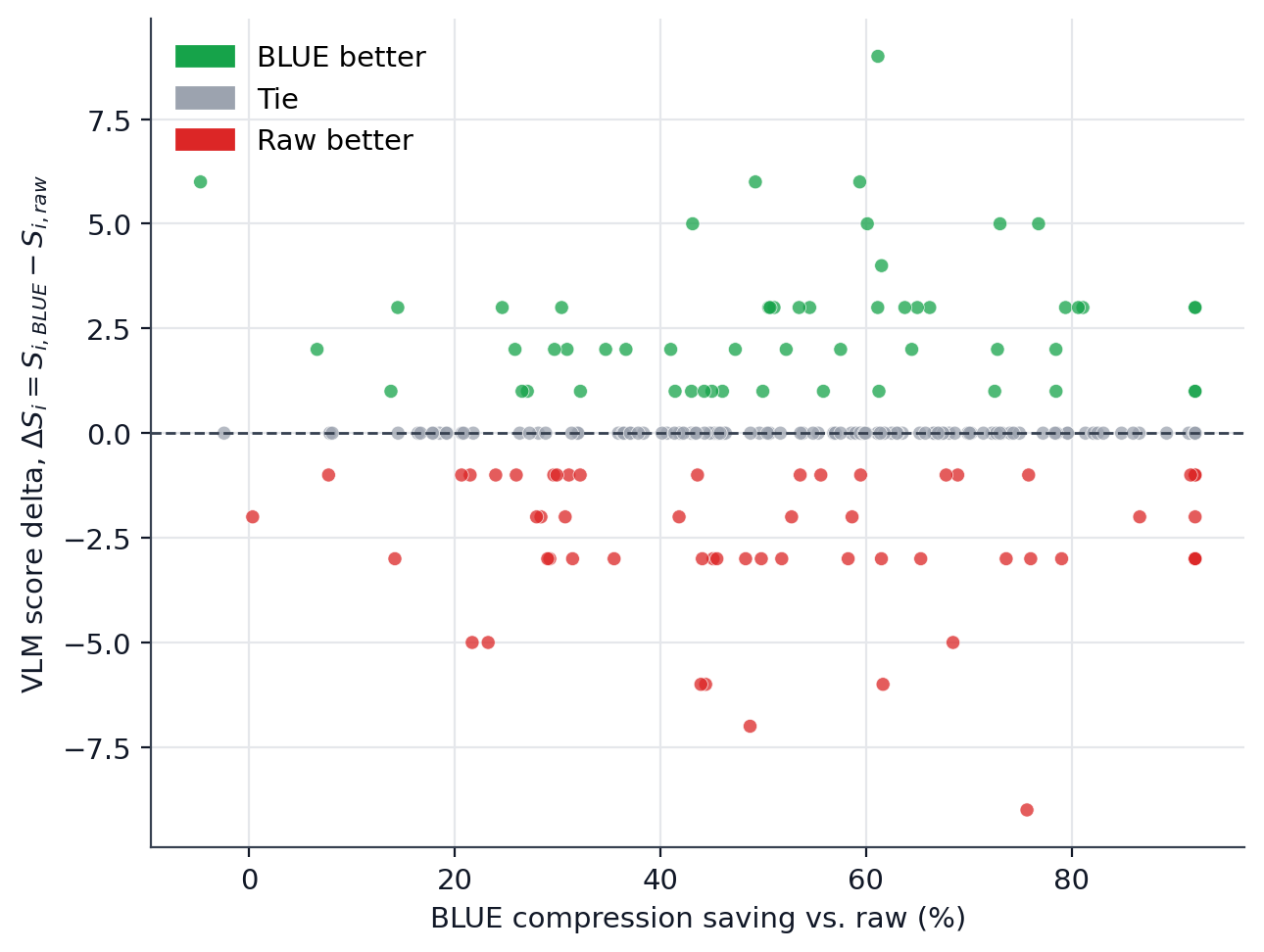}
\caption{Compression saving vs.\ VLM score delta on VIRAT ($n = 227$ samples). Each point is one
paired sample: $x$ = BLUE compression saving relative to raw H.265 (\%); $y$ = score delta, $\Delta
S_i = S_{i,BLUE} - S_{i,raw}$. Points are colored green/grey/red by whether BLUE won/tied/lost that
sample. The near-zero correlation ($r = 0.004$) and absence of a trend indicate that saving and
quality change are independent.}
\label{fig:4}
\end{figure}

A key question is whether higher compression predicts semantic degradation. On VIRAT, the Pearson
correlation between compression saving and VLM score change is $r = 0.004$, effectively zero. Fig.
\ref{fig:4} plots compression saving against score delta and shows no visible trend, confirming
that clips with higher BLUE compression savings do not show systematically worse VLM scores. This
supports the interpretation that BLUE mainly removes redundant static-background information rather
than foreground semantic evidence.

This negligible dependence of VLM quality on compression saving is consistent with both model
architecture and prior compression-robustness studies. Lossy codecs primarily discard
high-frequency detail and texture components that strongly affect human perception, but these
components are partly suppressed before VLM reasoning. Modern vision-language pipelines resize
frames, divide them into visual tokens, and pass them through transformer-based vision encoders
whose attention mechanisms tend to emphasize lower-frequency, global, and shape-level structure
rather than fine pixel-level artifacts \cite{ref18,ref19}. Prior work on compression and deep
vision models also reports that task performance is often stable under mild-to-moderate JPEG or
H.264 compression and degrades nonlinearly mainly under more severe compression regimes
\cite{ref20,ref21}. This is aligned with the surveillance-specific findings of O'Byrne et al.\
\cite{ref10}, where object-detection accuracy remained robust across substantial H.264 compression
before degrading at more aggressive settings. In VLMs, the decoupling is further reinforced
because the visual token budget is governed primarily by resolution and model tokenization policy
rather than encoded file size or pixel entropy \cite{ref22}. Therefore, the near-zero correlation
observed in Fig. \ref{fig:4} is expected: BLUE reduces redundant static-background information and
compression-sensitive high-frequency detail, while preserving the foreground activity and global
scene structure that dominate VLM semantic inference.

\subsection{CHAD Human-Anomaly Results}

CHAD provides a complementary test on human-centric anomalies. The category-level behavior
summarized in Table \ref{tab:chad-advantage} is more varied than the overall mean, indicating that
BLUE effects depend partly on action type and scene structure.

\begin{table}[H]
\centering
\caption{CHAD category-level BLUE advantage.}
\label{tab:chad-advantage}
\resizebox{\linewidth}{!}{%
\begin{tabular}{llll}
\toprule
\textbf{CHAD category} & \textbf{Direction} & \textbf{Approx. $\Delta$} & \textbf{Interpretation} \\
\midrule
Person lying on floor & Positive & +1.0 & BLUE improves detection/description \\
Playing & Positive & +1.8 & Strong improvement; small $n$ \\
Robbery & Positive & +0.7 & BLUE improves anomaly description \\
Throwing & Positive & +0.6 & BLUE improves action description \\
Running away & Near neutral & -0.1 & Essentially unchanged \\
Fight & Negative & -1.1 & BLUE lower than raw \\
Riding a bicycle & Negative & -1.4 & BLUE lower than raw \\
Normal & Negative & -1.7 & BLUE lower than raw \\
\bottomrule
\end{tabular}}
\end{table}

\begin{figure}[H]
\centering
\includegraphics[width=\linewidth]{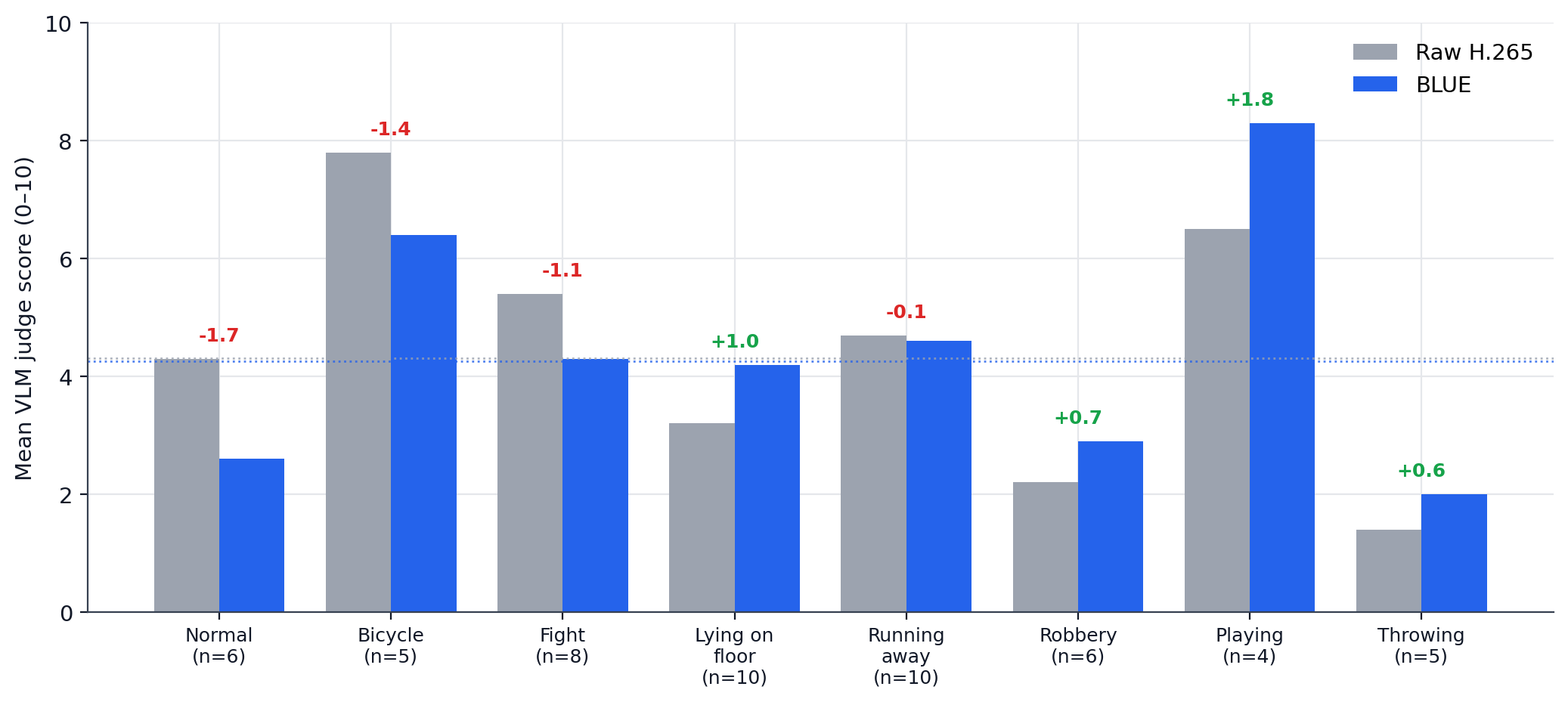}
\caption{Mean CHAD score by anomaly category, raw H.265 vs.\ BLUE. $n$ (below each category) is the
number of clips in that category, out of 54 total. Dotted lines mark the overall dataset mean per
variant (4.31 raw, 4.26 BLUE); $\Delta$ (above each bar pair) = mean(BLUE) $-$ mean(raw).}
\label{fig:5}
\end{figure}

\begin{figure}[H]
\centering
\includegraphics[width=0.85\linewidth]{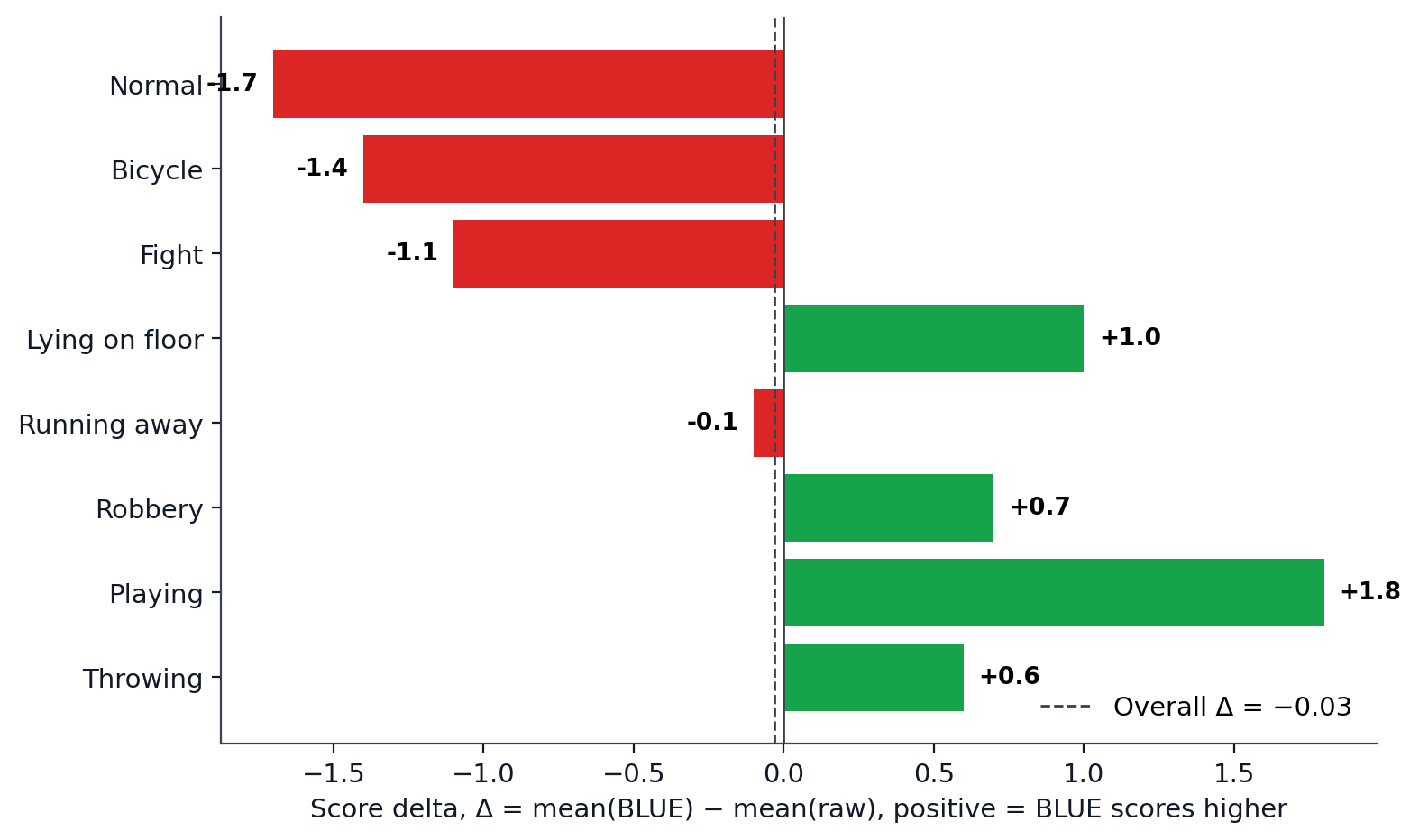}
\caption{BLUE score advantage by CHAD category, where $\Delta$ = mean(BLUE) $-$ mean(raw) and
positive values indicate BLUE scores higher than raw H.265. The dashed line marks the overall
dataset delta (-0.06).}
\label{fig:6}
\end{figure}

As shown in Fig. \ref{fig:5}, BLUE and raw H.265 remain broadly comparable across CHAD categories,
although some category-level differences are visible. Fig. \ref{fig:6} makes these differences
explicit by plotting BLUE's score advantage by category. BLUE improves several sparse foreground
anomaly categories, including person lying on floor, playing, robbery, and throwing. These cases
involve localized human activity against a relatively static background. BLUE underperforms on
normal, bicycle, and fight clips, which suggests that category-level effects should be interpreted
cautiously, particularly where sample counts are small.

The stronger conclusion from Table \ref{tab:chad-advantage} and Figs. \ref{fig:5}--\ref{fig:6} is
that BLUE does not degrade VLM inference on average, while category-level variation motivates
larger balanced evaluations.

\subsection{P-Frame-Based VLM Sampling Efficiency}

The CHAD experiment also evaluates whether BLUE can reduce the number of VLM calls in a continuous
surveillance pipeline. Although file-size reduction lowers storage and transmission cost, the
operational cost of VLM analytics is dominated by the number of frames sent for inference. Table
\ref{tab:pframe} summarizes the P-frame results and shows that BLUE creates a bitstream-level signal
for frame skipping by turning static-background frames into skip-heavy P-frames.

\begin{table}[H]
\centering
\caption{CHAD P-frame sampling efficiency.}
\label{tab:pframe}
\begin{tabular}{lccc}
\toprule
\textbf{Metric} & \textbf{Raw H.265} & \textbf{BLUE} & \textbf{Effect} \\
\midrule
Mean P-frame size & 53.1 KB & 16.3 KB & 3.26$\times$ smaller \\
Skip-heavy P-frames ($<$5 KB) & 1.4\% & 53.2\% & 38$\times$ more \\
Est.\ frames requiring VLM call & $\sim$99\% & $\sim$47\% & $\sim$53\% reduction \\
\bottomrule
\end{tabular}
\end{table}

\begin{figure}[H]
\centering
\includegraphics[width=\linewidth]{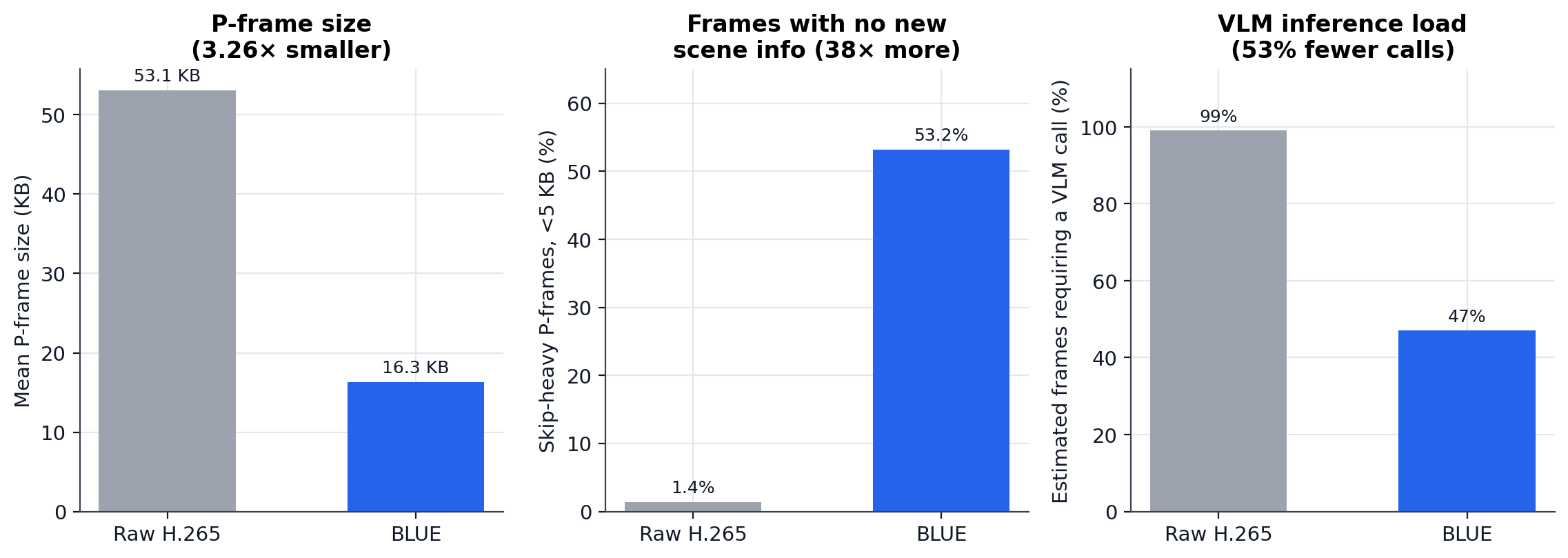}
\caption{Estimated VLM calls per CHAD clip for raw H.265 and BLUE ($n = 54$ clips), derived from
Table \ref{tab:pframe}. Left: mean P-frame size (KB). Center: share of P-frames below the 5 KB
skip-heavy threshold. Right: estimated share of frames that would require a VLM call under a
packet-size-based sampler.}
\label{fig:7}
\end{figure}

Under raw H.265, only 1.4\% of P-frames fall below the 5 KB skip-heavy threshold, so a
packet-size sampler would still process nearly every frame. Under BLUE, 53.2\% of P-frames fall
below this threshold. Fig. \ref{fig:7} shows the resulting reduction in estimated VLM calls per
CHAD clip.

\begin{figure}[H]
\centering
\includegraphics[width=0.85\linewidth]{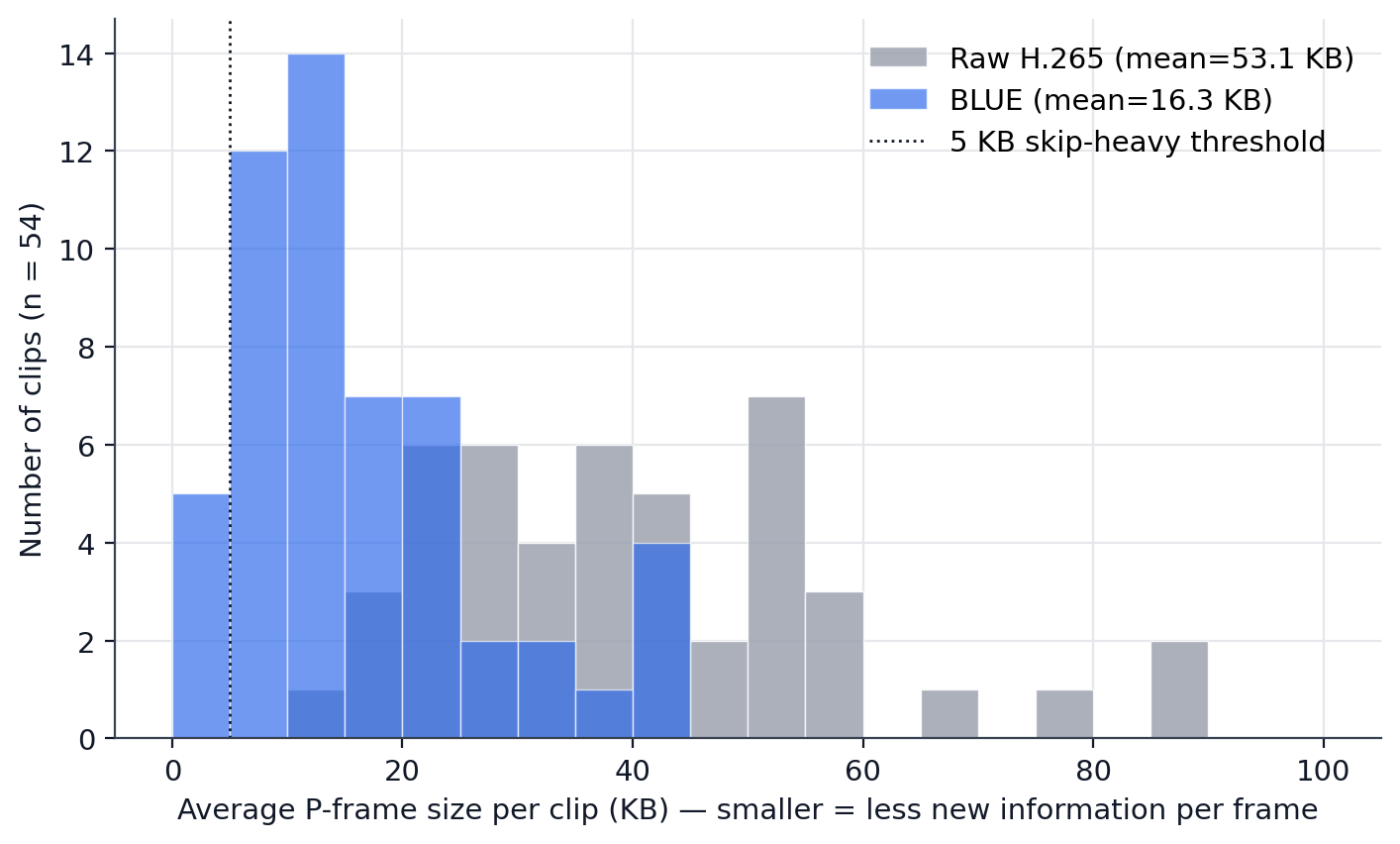}
\caption{Distribution of P-frame sizes for raw H.265 and BLUE, one value per clip ($n = 54$). The
dotted line marks the 5 KB skip-heavy threshold used throughout this paper.}
\label{fig:8}
\end{figure}

\begin{figure}[H]
\centering
\includegraphics[width=\linewidth]{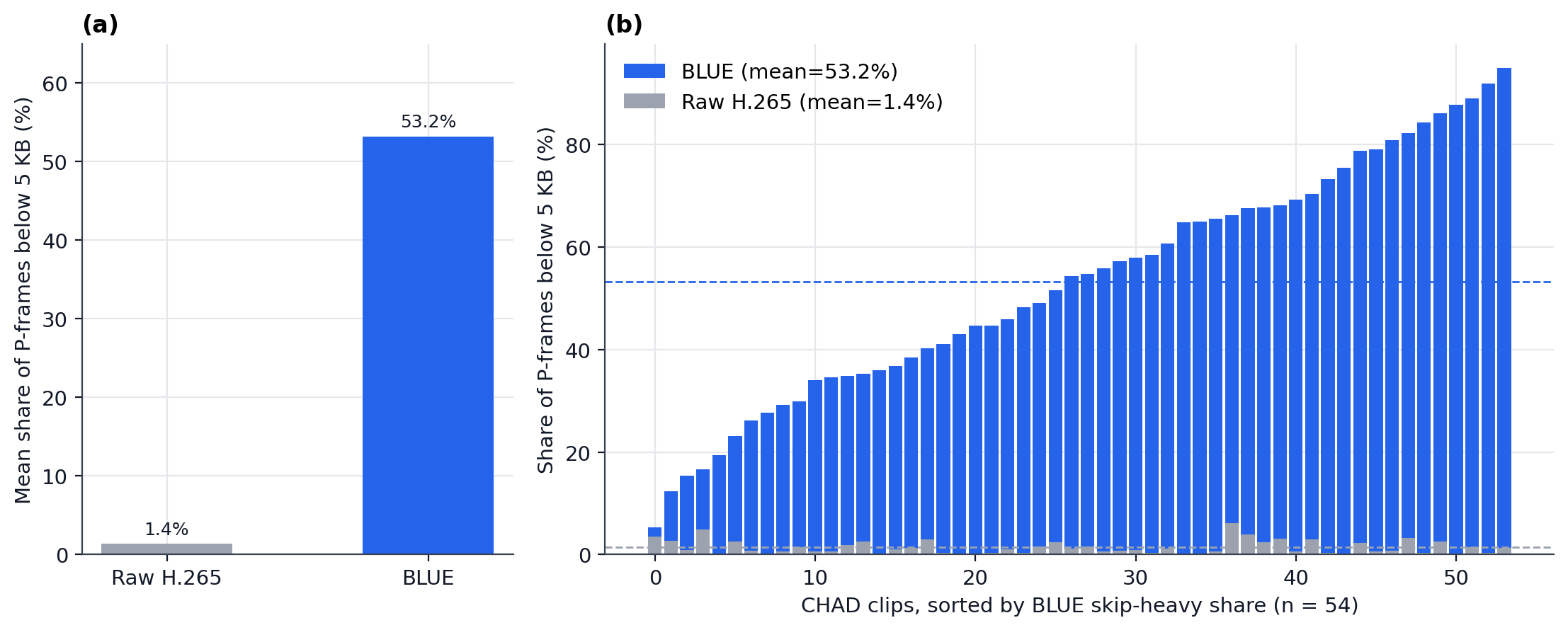}
\caption{Share of skip-heavy P-frames ($<$5 KB) for raw H.265 and BLUE. (a) Averaged across all 54
CHAD clips. (b) Per-clip share, clips sorted by BLUE skip-heavy share ($n = 54$); dashed lines mark
each variant's dataset mean.}
\label{fig:9}
\end{figure}

Fig. \ref{fig:8} shows that BLUE shifts the P-frame size distribution toward smaller packet sizes,
and Fig. \ref{fig:9}(a)--(b) show the corresponding increase in skip-heavy frames, both on average
and per clip. Since packet sizes can be read without decoding every frame, a VLM pipeline can skip
low-information frames before invoking the model.

\begin{figure}[H]
\centering
\includegraphics[width=\linewidth]{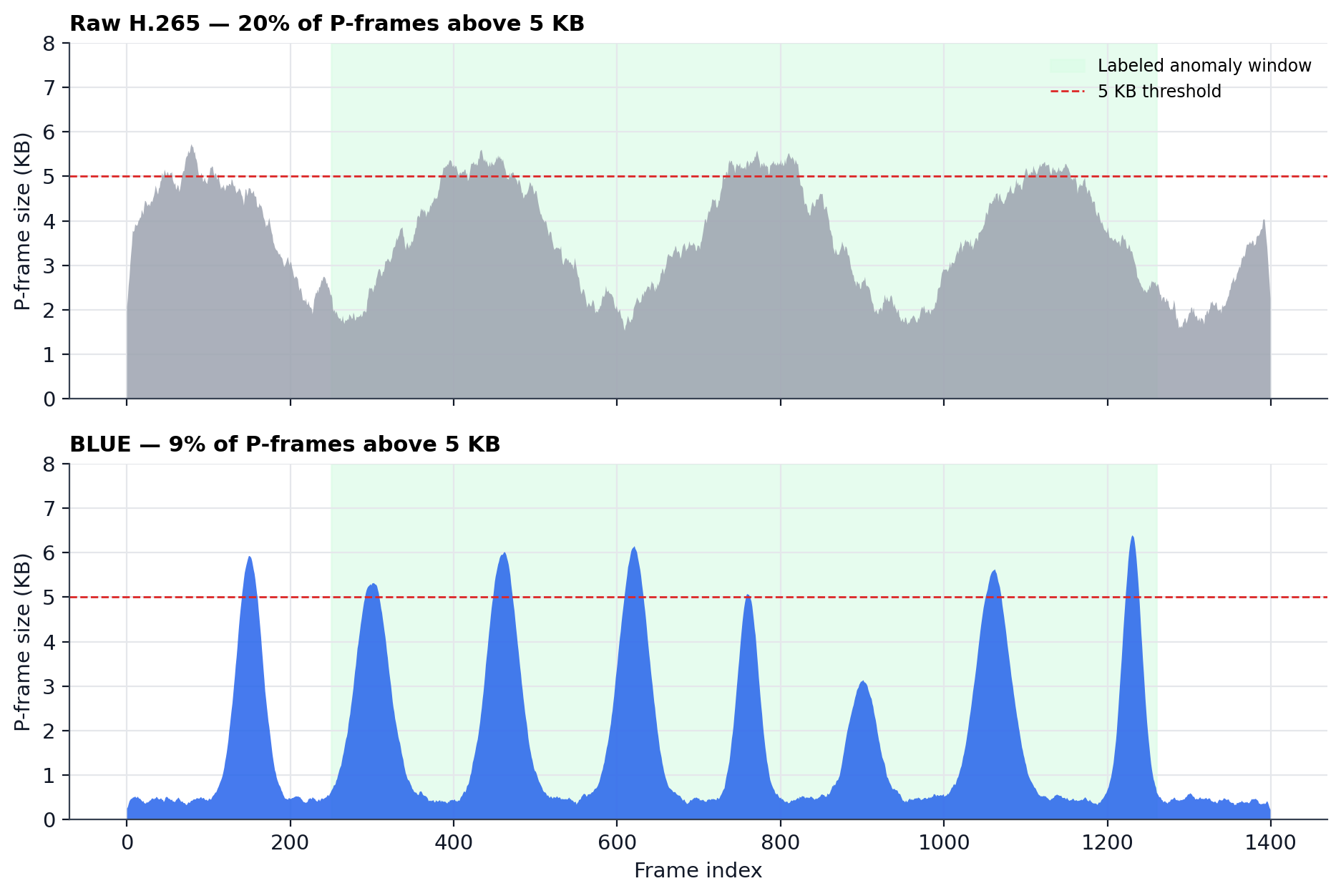}
\caption{P-frame activity profile for CHAD fight clip 2\_065\_1, raw H.265 (top) vs.\ BLUE (bottom).
The green band marks the labeled anomaly window; the dashed red line marks the 5 KB skip-heavy
threshold. Raw H.265 stays above the threshold for 20\% of frames throughout the clip, while BLUE
stays near zero outside motion events and exceeds the threshold for only 9\% of frames,
concentrated inside the anomaly window.}
\label{fig:10}
\end{figure}

Fig. \ref{fig:10} illustrates the mechanism on a representative CHAD fight clip: BLUE maintains low
P-frame cost outside major motion periods and spikes during real activity, enabling
packet-size-based selection of informative frames.

Overall, the results support three conclusions. First, BLUE preserves VLM semantic inference
quality under substantial compression. Second, compression saving is not predictive of semantic
quality loss, suggesting that BLUE suppresses redundant static-background information rather than
foreground semantic content. Third, BLUE enables a downstream inference-efficiency mechanism by
exposing low-information frames through P-frame size. These properties make BLUE a machine-centric
compression layer for scalable VLM-based surveillance analytics.

\section{Conclusion}

This study examined whether BLUE compression affects vision-language model inference on
surveillance video. Using paired raw H.265 and BLUE-compressed H.265 samples from VIRAT and CHAD,
the evaluation shows that BLUE achieves substantial compression without measurable loss in semantic
inference quality. On VIRAT, the mean VLM score remained effectively unchanged, with a difference
of approximately -0.01 on a 0-10 scale across 227 paired event samples. On CHAD, the mean scores
were also near-equivalent, 4.31 for raw H.265 and 4.26 for BLUE, across 54 clips. The balanced
win/loss distribution further indicates that BLUE does not introduce a systematic degradation in
VLM-based event or anomaly understanding.

The results also show that BLUE's compression benefit is not obtained at the expense of machine
perception. In the VIRAT experiment, compression saving and VLM score change were essentially
uncorrelated ($r = 0.004$), indicating that higher compression did not predict semantic quality
loss. This supports the interpretation that BLUE primarily removes redundant static-background
information while preserving foreground activity relevant to VLM inference.

Beyond file-size reduction, BLUE provides an additional operational advantage for VLM-based video
analytics. On CHAD, BLUE increased the proportion of skip-heavy P-frames from 1.4\% to 53.2\%,
reducing the estimated number of required VLM calls by approximately 53\%. Since this decision can
be made from compressed packet metadata without full frame decoding, BLUE can serve as both a
compression layer and a low-cost frame-selection mechanism.

Overall, the findings suggest that BLUE is a machine-centric compression approach suitable for
fixed-camera surveillance analytics. It reduces storage and transmission requirements, preserves
VLM semantic performance, and enables more efficient inference scheduling. Future work should
extend the evaluation to larger and more diverse datasets, multi-frame VLM reasoning, and real-time
deployments where event-start sampling and temporal aggregation can further reduce transient
single-frame failure cases.


\begin{thebibliography}{99}

\bibitem{ref1} G. J. Sullivan, J.-R. Ohm, W.-J. Han, and T. Wiegand, ``Overview of the High
Efficiency Video Coding (HEVC) standard,'' \textit{IEEE Transactions on Circuits and Systems for
Video Technology}, vol. 22, no. 12, pp. 1649--1668, Dec. 2012.

\bibitem{ref2} B. Bross, Y.-K. Wang, Y. Ye, S. Liu, J. Chen, G. J. Sullivan, and J.-R. Ohm,
``Overview of the Versatile Video Coding (VVC) standard and its applications,'' \textit{IEEE
Transactions on Circuits and Systems for Video Technology}, vol. 31, no. 10, pp. 3736--3764, Oct.
2021.

\bibitem{ref3} J. Han et al., ``A technical overview of AV1,'' \textit{Proceedings of the IEEE},
vol. 109, no. 9, pp. 1435--1462, Sept. 2021.

\bibitem{ref4} S. Ma, X. Zhang, C. Jia, Z. Zhao, S. Wang, and S. Wang, ``Image and video compression
with neural networks: A review,'' \textit{IEEE Transactions on Circuits and Systems for Video
Technology}, vol. 30, no. 6, pp. 1683--1698, Jun. 2020.

\bibitem{ref5} D. Ding, Z. Ma, D. Chen, Q. Chen, Z. Liu, and F. Zhu, ``Advances in video compression
system using deep neural network: A review and case studies,'' \textit{Proceedings of the IEEE},
vol. 109, no. 9, pp. 1494--1520, Sept. 2021.

\bibitem{ref6} L.-Y. Duan, J. Liu, W. Yang, T. Huang, and W. Gao, ``Video coding for machines: A
paradigm of collaborative compression and intelligent analytics,'' \textit{IEEE Transactions on
Image Processing}, vol. 29, pp. 8680--8695, 2020.

\bibitem{ref7} W. Gao, S. Liu, X. Xu, M. Rafie, Y. Zhang, and I. Curcio, ``Recent standard
development activities on video coding for machines,'' \textit{arXiv preprint arXiv:2105.12653},
2021.

\bibitem{ref8} M. Aqqa, P. Mantini, and S. K. Shah, ``Understanding how video quality affects
object detection algorithms,'' in \textit{Proc. International Conference on Computer Vision Theory
and Applications (VISAPP)}, 2019, pp. 96--104.

\bibitem{ref9} K. Kajak, ``Impact of video compression on the performance of object detection
algorithms in automotive applications,'' M.S. thesis, KTH Royal Institute of Technology, Stockholm,
Sweden, 2020.

\bibitem{ref10} M. O'Byrne, Vibhoothi, M. Sugrue, and A. Kokaram, ``Impact of video compression on
the performance of object detection systems for surveillance applications,'' \textit{arXiv preprint
arXiv:2211.05805}, 2022.

\bibitem{ref11} T. Tanaka, A. Harell, and I. V. Bajic, ``Does video compression impact tracking
accuracy?'' in \textit{Proc. IEEE International Symposium on Circuits and Systems (ISCAS)}, 2022.

\bibitem{ref12} K. Shao et al., ``When tokens talk too much: A survey of multimodal long-context
token compression across images, videos, and audios,'' \textit{arXiv preprint arXiv:2507.20198},
2025.

\bibitem{ref13} J. Fei et al., ``Small vision-language models are smart compressors for long video
understanding,'' \textit{arXiv preprint arXiv:2604.08120}, 2026.

\bibitem{ref14} S. Oh et al., ``A large-scale benchmark dataset for event recognition in
surveillance video,'' in \textit{Proc. IEEE Conference on Computer Vision and Pattern Recognition
(CVPR)}, 2011, pp. 3153--3160.

\bibitem{ref15} A. Danesh Pazho, G. Alinezhad Noghre, B. Rahimi Ardabili, C. Neff, and H. Tabkhi,
``CHAD: Charlotte Anomaly Dataset,'' in \textit{Proc. Scandinavian Conference on Image Analysis
(SCIA)}, 2023, pp. 50--66.

\bibitem{ref16} W. Kwon et al., ``Efficient memory management for large language model serving with
PagedAttention,'' in \textit{Proc. ACM Symposium on Operating Systems Principles (SOSP)}, 2023, pp.
611--626.

\bibitem{ref17} FFmpeg Developers, ``ffprobe documentation,'' FFmpeg multimedia framework
documentation. [Online]. Available: \url{https://ffmpeg.org/ffprobe.html}. Accessed: Jul. 18,
2026.

\bibitem{ref18} N. Park and S. Kim, ``How do vision transformers work?'' in \textit{Proc.
International Conference on Learning Representations (ICLR)}, 2022.

\bibitem{ref19} M. Naseer, K. Ranasinghe, S. Khan, M. Hayat, F. S. Khan, and M.-H. Yang,
``Intriguing properties of vision transformers,'' in \textit{Proc. Advances in Neural Information
Processing Systems (NeurIPS)}, 2021.

\bibitem{ref20} S. Dodge and L. Karam, ``Understanding how image quality affects deep neural
networks,'' in \textit{Proc. Eighth International Conference on Quality of Multimedia Experience
(QoMEX)}, 2016.

\bibitem{ref21} M. Poyser, A. Atapour-Abarghouei, and T. P. Breckon, ``On the impact of lossy image
and video compression on the performance of deep convolutional neural network architectures,'' in
\textit{Proc. International Conference on Pattern Recognition (ICPR)}, 2020, pp. 2830--2837.

\bibitem{ref22} P. Wang et al., ``Qwen2-VL: Enhancing vision-language model's perception of the
world at any resolution,'' \textit{arXiv preprint arXiv:2409.12191}, 2024.

\end{thebibliography}
\end{document}